# The Atacama Large Millimeter/submillimeter Array (ALMA) Band-1 Receiver

Yau De(Ted) Huang [a*], Oscar Morata [a], Patrick Michel Koch [a], Ciska Kemper [a], Yuh-Jing Hwang [a], Chau-Ching Chiong [a], Paul Ho [a], You-Hua Chu [a], Chi-Den Huang [b], Ching-Tang Liu [b], Fang-Chia Hsieh [b], Yen-Hsiang Tseng [b], Shou-Hsien Weng [a], Chin-Ting Ho [a], Po-Han Chiang [a], Hsiao-Ling Wu [a], Chih-Cheng Chang [a], Shou-Ting Jian [a], Chien-Feng Lee [a], Yi-Wei Lee [a], Satoru Iguchi [c], Shin'ichiro Asayama [c], Daisuke Iono [c], Alvaro Gonzalez [c], John Effland [d], Kamaljeet Saini [d], Marian Pospieszalski [d], Doug Henke [e], Keith Yeung [e], Ricardo Finger [f], Valeria Tapia [f], Nicolas Reyes [f]

[a] Academia Sinica Institute of Astronomy and Astrophysics, Taipei, Taiwan, ROC.
[b] Aeronautical System Research Division, National Chung-Shang Institute of Science and Technology, Taichung, Taiwan, ROC.
[c] National Astronomical Observatory of Japan, Tokyo, Japan (NAOJ)
[d] National Radio Astronomy Observatory, Charlottesville, VA, USA
[e] NRC-CNRC Herzberg Institute of Astrophysics, Victoria, BC, Canada
[f] Universidad de Chile, Santiago, Chile

**ABSTRACT**

The Atacama Large Millimeter/submillimeter Array(ALMA) Band 1 receiver covers the 35-50 GHz frequency band. Development of prototype receivers, including the key components and subsystems has been completed and two sets of prototype receivers were fully tested. We will provide an overview of the ALMA Band 1 science goals, and its requirements and design for use on the ALMA. The receiver development status will also be discussed and the infrastructure, integration, evaluation of fully-assembled band 1 receiver system will be covered. Finally, a discussion of the technical and management challenges encountered will be presented.

## 1.    INTRODUCTION

The Academia Sinica Institute of Astronomy and Astrophysics (ASIAA) is leading the construction of the Band 1 (35-50 GHz) receiver system to the ALMA project, in collaboration with the National Research Council of Canada Herzberg Astronomy and Astrophysics (NRC Herzberg), the National Radio Astronomy Observatory (NRAO) in USA, the University of Chile (UCh) in Chile, and the National Astronomical Observatory of Japan (NAOJ). The main scientific goal of ALMA Band 1 is to provide access to frequencies ~40 GHz at high resolution and sensitivity from the southern hemisphere. The publicly available ALMA Band 1 Science Case [1] contains an extensive and detailed description of numerous scientifically compelling cases proposed for this new receiver. ALMA Band 1 will be able to study a broad range of astrophysical environments: from solar studies, nearby stars, and molecules in nearby molecular clouds to distant galaxy clusters and the re-ionization edge of the Universe. The ALMA Band 1 Science Team redefined the Band 1 frequency range to cover the 35-50 GHz band, with an extension to 51 GHz. This shift allows the observation of some important (CS 1-0) and/or interesting (HDO) lines, it increases the available frequencies by a factor of ~10%, it slightly improves the angular resolution of continuum observations, and it better exploits the advantages of the dry ALMA site. Band 1 will allow ALMA to bridge the gap between the mm/sub-mm and cm radio astronomy and, given the large ratio of the available bandwidth to frequency, it will be able to study a large range of energy regimes, from quiescent gas to maser emission, solar flares, and Active Galactic Nuclei.

The plan is to complete and deliver all 73 receiver units by the end of 2019. Not only are the technical requirements for the receivers far more stringent than any existing receiver system at this frequency band, but also the development and delivery schedule is challenging. To meet this requirement, the integration/test infrastructure and human resources must be planned to sustain the requirements over the next several years. Industrial involvement is also one of the important elements in the project planning. This document will present an overview of the Band 1 receiver project, the status of receiver development, and the progress of establishing the infrastructure and the projected project time-line.

## 2. SCIENCE GOAL

- The two main scientific cases of ALMA Band 1 are also two of the ALMA Level One Science Goals:
- *the study of grains in protoplanetary disks to sizes as large as ~1 cm.* Planet formation takes place in disks of dust and gas surrounding young stars. It is in protoplanetary disks where dust grains must agglomerate from sub-micron sizes to larger pebbles, rocks and planetesimals, on a timescale of tens of Myr in order to form terrestrial-like planets. The identification of *where* and *when* dust coagulation occurs is critical to constrain current models of planetary formation. The growth from sub-micron to micron-sized particles can be traced with IR spectroscopy and imaging polarimetry. But, with Band 1, ALMA will be the only facility able to follow the evolution of dust grain growth from mm-sized to cm-sized pebbles in protoplanetary disks. Thus, Band 1 has a crucial role in the localization and characterization of the coagulation processes of dust grains to centimeter sizes in protoplanetary disks.
- *the detection of molecular gas from nearby to high-redshift galaxies ($6 < z < 10$ ).* In order to study the first generation of galaxies, it is necessary to study the star-formation properties of high-redshift galaxies. The detection of several transitions of CO, and also of dust continuum, in some high-$z$ sources show that there was already a significant abundance of metals and dust by those early epochs, but current sensitivity limitations restrict these findings to the small fraction of hyperluminous infrared galaxies. Band 1 will follow-up on high-$z$ detections done in southern sources and will be able to detect rotational transitions of CO, from 1--0 to 6--5 depending on the redshift (for $z$<11). In particular, Band 1 will be able to detect the CO 3--2 line at redshifts of the era of reionization in the range $6 < z < 9$ (other ALMA bands can do so with higher-$J$ lines, which may be less excited). Additionally, Band 1 will be able to cover redshifts $1.3 < z < 9$, with a few gaps, using the CO 1--0, 2--1, and 3--2 lines. Finally, the wide Band 1 frequency range will open the possibility of carrying out ``blank-sky'' surveys pointing ALMA towards one location and stepping through the whole Band 1 frequency range.
- But, the scientific case for ALMA Band 1 is much more extended:
- study of the fine structure of chemical differentiation in cloud cores, using molecular transitions close to the fundamental rotational levels, which will probe the smallest length scales of chemical variation in cloud cores
- studies of Very Small Grains and spinning dust: the somewhat elusive emission, spatially correlated with thermal dust emission and peaking at ~40 GHz, that traces extremely small dust grains, which may play a central role in the chemical and thermal balance of the ISM.
- observations of complex carbon-chain molecules, including the amino-acids and sugars from which life on Earth may have originally evolved
- study of the bulk of the gas in the Universe, the cold molecular ISM, through the observation of many molecular lines, such as SiO, CS, methanol, carbon-chains, etc, which will allow the study the star formation processes in nearby galaxies in a similar way as it is routinely done for Galactic studies
- the study of the Sunyaev-Zel'dovich effect to probe the physics of galaxy clusters performing high-resolution and high-sensitivity observations of SZE clusters with the goal of detecting cluster substructures
- Solar observations: Band 1 will be the best band to simultaneously observe Solar flares and their acceleration sites in the Sun
- study of magnetic fields through the use of the Zeeman effect: Band 1 receivers will provide the opportunity to measure the initial mass-to-flux ratio of molecular cores through the detection of the Zeeman effect in spectral lines

# 3. DESIGN

The ALMA Band 1 Receiver will formally operate in the 35-50 GHz frequency band and coverage in the 50 – 52 GHz range is to be achieved on a best-effort basis. The formal frequency range is 35% bandwidth with a center frequency of 42.5 GHz. The receiver comprises of three subsystems: optics, cold cartridge and warm cartridge. The technology used for Band 1 is dual-polarization, SSB heterodyne receiver covering the specified frequency range with an IF band of 4-12 GHz.

A block diagram of the receiver is shown in Figure 1. Unlike the higher frequency ALMA cartridge bands from Band-3 to Band-10 where superconductor-insulator-superconductor (SIS) mixers were used to achieve required sensitivity for signal detection, Band 1 uses hetero-junction field effect transistor (HFET) cryogenically cooled amplifiers as the first receiver stage. In the Band 1 cartridge design, despite the unavoidable optics loss and insertion loss of the orthomode transducer (OMT)[2], the sensitivity of the receiver is mainly determined by the noise and gain performance of the cryogenic low noise amplifiers (LNAs). The incoming beam from the ALMA antenna is refocused by a warm lens, which acts as vacuum window, and collected by a corrugated feedhorn at the 15K stage. The two orthogonal polarizations (0 and 1) are then split using an OMT[2]. The signals are then amplified with cryogenic low noise amplifiers fixed to the 15K stage [3]. Since these amplifiers are optimized for low noise, bandwidth and gain rather than input/output RF impedance match, the use of Q Band isolators at the cold cartridge 300K plate section minimizes ripple from signal reflections between the cryogenic amplifier and room temperature amplifier unit.

Since the first stage consists of LNAs, room temperature down-conversion is used in the Band 1 system without compromising receiver sensitivity. To minimize the impact of mixer noise to the system, a commercially available Q-band amplifier is used prior to the mixer to further amplify the signal. The upper sideband configuration is realized by using high-pass filter [4] after the room temperature Q-band amplifier. The commercially available mixer was chosen to provide high reliability and to simplify maintenance. The IF signals from the mixer are then input to a commercially available IF amplifier to meet ALMA IF power level specifications.

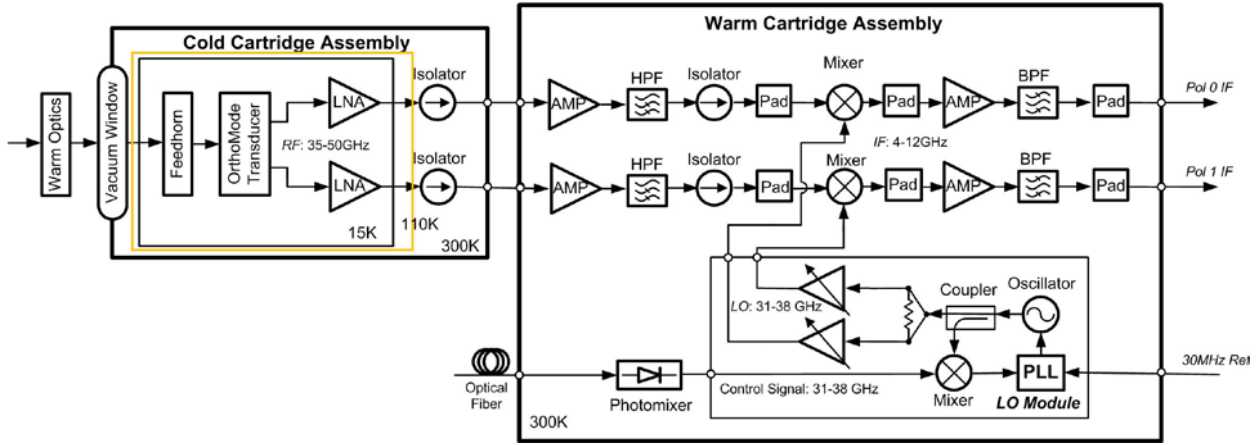

Figure 1: Block Diagram of the Band 1 cartridge assembly

# 4. SYSTEM PERFORMACE

### 4.1 Noise temperature and IF power performance

Noise temperature and IF power performance Table 1 summarizes the key specifications for Band 1, based on the requirements set by the ALMA Scientific Advisory Committee[5][6]. In this section we summarize the performance achieved for the prototype Band 1 receiver cartridges that were measured in a cartridge test cryostat. The receiver noise temperature was measured by a standard Y-Factor technique using 300K and 80K black body radiators presented in front of the receiver. The receiver noise temperature, integrated over the 4-12 GHz IF band, is presented in Figure 2 and Figure 3 as a function of LO frequency. The measured receiver noise temperature of the Band 1 receiver cartridge is below 30K full-band for Pol-0 and Pol-1, except some resonant spikes reach 31K.

Table-1 Band 1 CARTRIDGE MAIN PERFORMANCE REQUIREMENTS

| Parameter | Specification |
| --- | --- |
| RF port frequency range | 35.0-50.0GHz/50.0-52.0(Best Effort) |
| LO Port Frequency rang | 31.0-40.0 GHz |
| IF Port Frequency range | 4.0 –12.0 GHz |
| SSB Noise Temperature | $\leq$ 25 K over 80% of band <br> $\leq$ 32 K over entire band |
| Image band suppression and sideband mismatch | $\geq$ 10 dB over 90% of IF frequency range. <br> > 7 dB over entire IF frequency range. |
| Large signal gain compression | 5% caused by the different RF load temperatures of 77K and 373K |
| The total power within the IF band | >-32dBm to -22dBm |
| IF power variations | 4dB p-p over any 2GHz window <br> 7dB p-p full band |
| large signal gain compression | <5% @load exchange between 77 K and 373K |
| Amplitude stability: Allan variance | $4.0 \times 10^{-7}$ for timescales in the range of $0.05\ s \leq T \leq 100\ s$ <br> $3.0 \times 10^{-6}$ for T = 300 seconds. |
| Signal path phase stability | 22fs over 300s |
| Aperture efficiency | >80% |

The pre-production cartridge maximum receiver IF output power variation within any 2 GHz wide window is shown in Figure 3 and Figure 4. and indicates system performance is not compliant with peak-to-peak variation of 4 dB in a 2 GHz window. This 2 GHz power variation is an important receiver parameter, because excessive power variation across the 2 GHz wide sampler bandwidth limits the dynamic range of the backend.  The ripple budget indicates the dominant components are the cold LNA and the mis-match between the cold LNA and warm Q-Band LNA. The IF chain, including the IF amplifier, is not the dominant component in ripple budget, as proven by analysis and measurements.  A more accurate receiver model will be obtained after collecting additional data using the pre-production cartridges. This would allow us to confirm the 2GHz window specification performance.

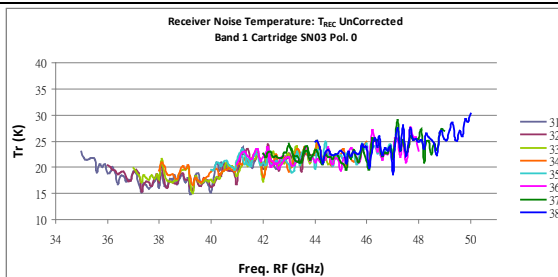

Figure 2: Measured receiver noise temperature for Pol 0

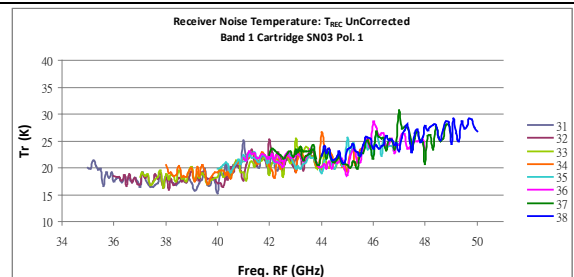

Figure 3: Measured receiver noise temperature for Pol 1.

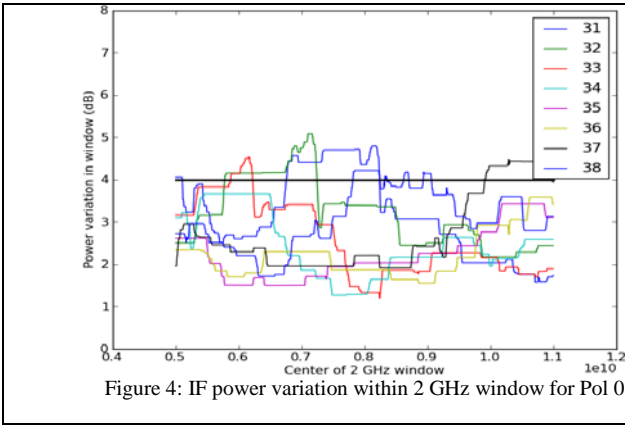
Figure 4: IF power variation within 2 GHz window for Pol 0

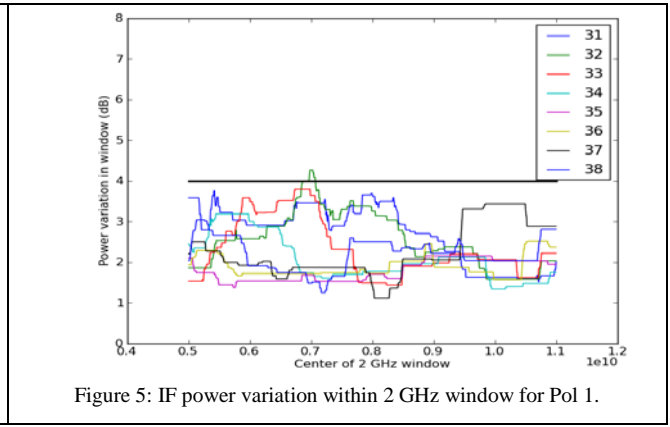
Figure 5: IF power variation within 2 GHz window for Pol 1.

## 4.2 Image band suppression and Gain compression

The image band suppression characterization method injects a weak RF signal from a signal generator into the receiver input and maintains the same input signal power for both USB and LSB. The measured result is mostly greater than 25dB suppression, except the case of LO = 38GHz and IF = 4GHz, where the corresponded lower sideband RF frequency is 34GHz and the image-band suppression is 10dB, but that is still compliant with specifications. Gain compression is measured by injecting an external signal in the input port of RF and measuring the IF output power from the IF output channels. The test is performed by sweeping the RF input power at different LO frequencies. The gain compression measurement result indicates the Band 1 receiver is compliant with the 5% specification.

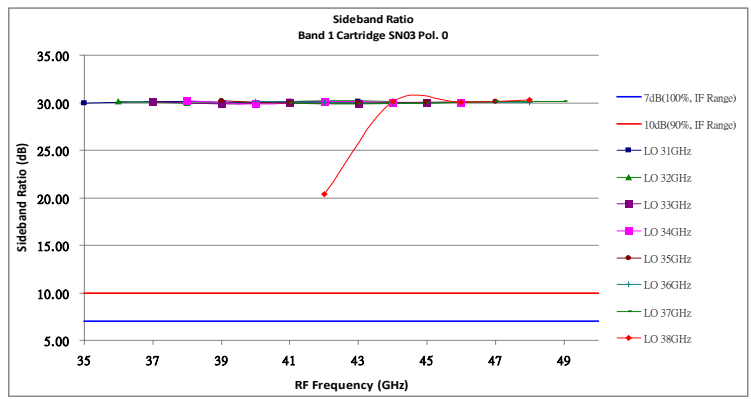

Figure 6: Measured image band suppression for Pol 0.

## 4.3 Amplitude stability and Phase stability

Amplitude stability and Phase stability Band 1 amplitude stability measures below $2.0 \times 10^{-7}$ for timescales in the range of $0.05 \text{ s} \leq T \leq 300 \text{ s}$, and that is compliant with the specification as shown for an LO frequency of 38 GHz in Figure 6 (Pol 0) and Figure 7 (Pol 1). Measured phase stability for all the LO frequencies is below 10fs, which meets specifications.

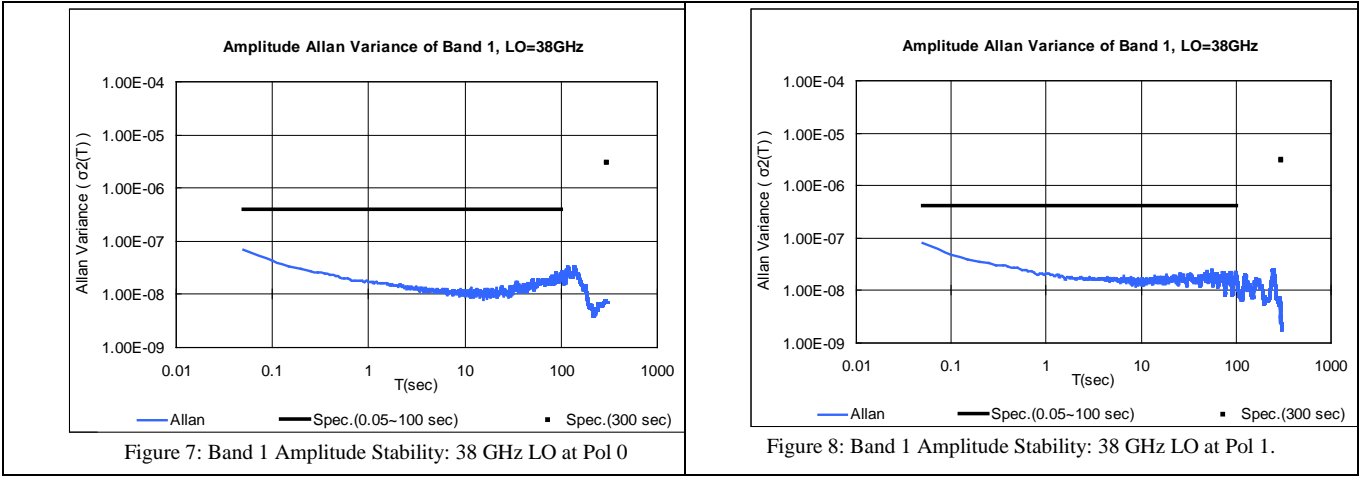

Figure 7: Band 1 Amplitude Stability: 38 GHz LO at Pol 0

Figure 8: Band 1 Amplitude Stability: 38 GHz LO at Pol 1.

### 4.4 Receiver Optics performance

The ALMA band 1 optics layout is shown in Figure 8. It consists of a compact spline-profile corrugated horn machined from a single block of aluminum, a low-loss HDPE one-zone bi-hyperbolic lens and the holder of the lens, including vacuum interfaces. The optics design of the ALMA band 1 receiver considers the beam should be tilted 2.48° with respect to the cryostat axis, in order to point towards the sub-reflector. In consequence, the horn and the lens are tilted toward the sub-reflector in the same angle. Moreover, in order to minimize the truncation in cryostat apertures, the horn has been placed 5 mm away from the 15K IR filter. This is as close as possible without risking collision during cool-down. The lens is located at 80 mm from the cryostat top plate on a holder structure. Since the lens is used as vacuum window, the assembly is composed of several components, a holder that provides the appropriate distance, a support ring that uniformly distributes the load across the lens in order to avoid vacuum leaks and O-rings as vacuum seals. The diameter of the lens and holder are as large as possible without risking collision with some instruments placed above the ALMA cryostat, in particular, the support arm of the Water Vapor Radiometer. This mechanical restriction on the lens size has an important impact on optical performance, since this effectively limits the maximum achievable aperture efficiency to 84%, which is close to the 80% ALMA specification. The optical design achieved values of aperture efficiency between 82 and 83.5%. Measured aperture efficiency has decreased 1.7% in average with respect to the simulations, showing 4 frequency points below 80% which is out of specifications. This is mostly due to effects associated with truncation at cryostat apertures, and reflection and impedance coupling in cryostat IR filters.

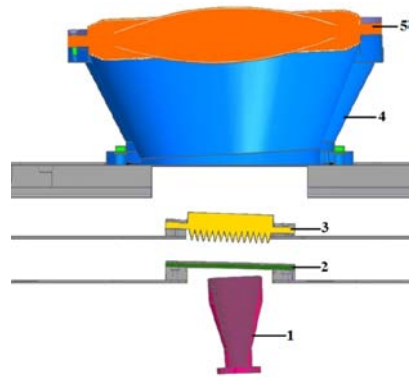

Figure 9: Cross-section of the baseline optical design. 1: Spline profile horn antenna, 2: 15 K infrared filter, 3: 110 K infrared filter, 4: lens holder, 5: One-zone modified Fresnel lens.

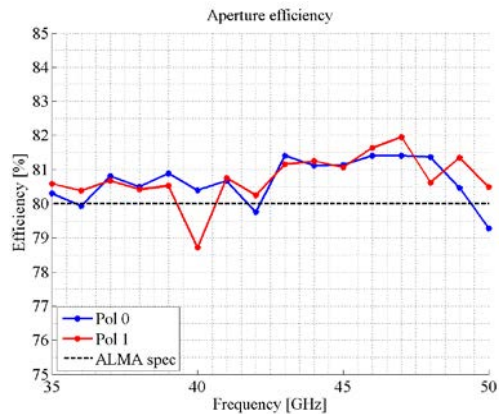
Figure 10: Full system efficiencies for polarizations 0 and 1.

# 5. BAND 1 RECEIVER PRODUCTION

Development and production of the ALMA Band 1 receiver benefits from the use of standard commercially obtainable components where possible, and only when no alternatives are available, custom made components are used. During the production phase, NRAO and NRC Herzberg will not only contribute components development but will also supply the production components: 1st LO (NRAO), Cold Low Noise Amplifier (NRAO) and OMT (NRC Herzberg) in the same manner as STFC/RAL (ALMA cartridge body). The University of Chile is focusing on supplying the optics components including the horn and lens. Band 1 receiver production will be carried out by the Aeronautical System Research Division (ASRD) who previously, in cooperation with ASIAA, staffed the East Asia Front End Integration Center (EA FEIC) that constructed and tested about a third of the 73 front ends during ALMA production.

## 5.1 Project Management Approach

The Band 1 Project Manager is responsible for managing and executing this project according to a formal ALMA Project Plan and its Subsidiary Management Plans. The project team consists of personnel from the ASIAA technical group, ASRD technical group, ASRD quality control/assurance group, and the testing group. The project manager closely interacts with all these resources to optimize project planning.

The project team is a matrix in that team members from each organization continue to report to their organizational functional management throughout the duration of the project. The project manager is responsible for communicating with organizational functional managers on the progress and performance of each project resource. The Band 1 Project Management framework embodies the project life cycle and five major project management process groups.

## 5.2 Facility and Staff

Band 1 cartridge integration and testing converted the former EA FEIC to the Band 1 integration and testing lab. The test system was converted and modified for Band 1 production unit integration and testing during the development phase (2013-2015) in conjunction with the assembly and testing of the prototype receivers, and the test procedures were also refined in the same period.

Figure 10 shows the Band 1 production process and relationship (outside partners and vendors are omitted from this diagram). Trained personnel is necessary for both tasks during the production phase, and the availability of qualified staff could adversely impact the Band 1 cartridge delivery schedule. To mitigate this risk, we are establishing a training program to ensure new staff rapidly acquire the necessary knowledge and a cross-training technique is being developed where multiple team members possess the same skill sets.

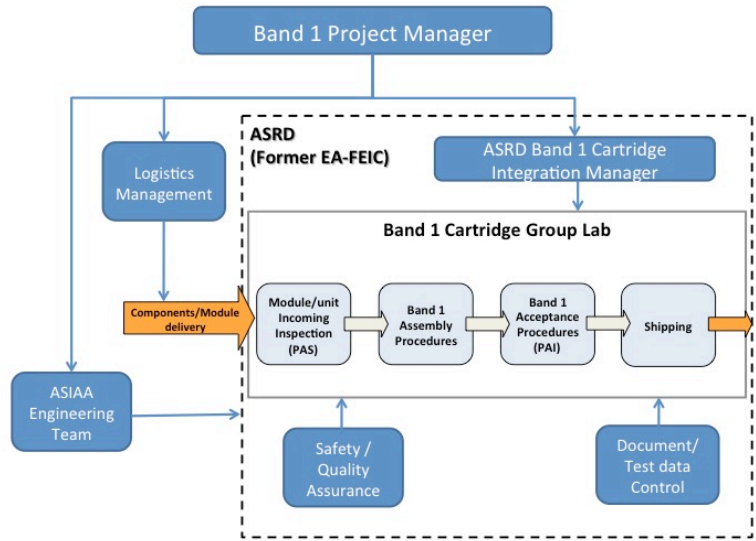

Figure 11: Production process

**5.3 Production Schedule**

The first year of the production phase in year 2016 is dedicated to setting up a second complete receiver test line that allows for a higher production rate of approximately two receivers per month starting from middle of 2017. The goal is to complete the last (73rd) receiver delivery to the ALMA Operation Support Facility (OSF) by the end of 2019.

# 6. CONCLUSION

We have assembled and fully characterized an ALMA band 1 receiver that exhibits good sensitivity and wide bandwidth coverage without performance degradation. The design is constrained by the truncation and coupling effects of the IR filters, which are already installed in the receivers and whose modification is prohibitively expensive, that degrade beam efficiencies. Despite that constraint, we are still able to demonstrate an optics design with at least 80% except for 4 frequencies. Achieving the aggressive schedule requires careful planning and cooperation among a matrix of widely-scattered teams, using the best available components at the time of assembly. The EA FEIC facility, whose teams successfully integrated and tested about one third of the 70 ALMA receivers, is leveraged for Band 1 production integration and testing to provide an experienced staff. Established PA/QA processes of that facility will be employed to continue and promote the safety and product assurance activities for Band 1 production.

# 7. ACKNOWLEDGMENT

The authors would like to thank Academia Sinica for support to establish the international collaboration and consortium on the ALMA band-1 development. U.Chile thanks the support of CONICYT through the fund CATA-Basal PFB06 for band-1 optics development.

# REFERENCE LINKING